\documentclass[11pt,a4paper]{article}
\def\bea{\begin{eqnarray}}
\def\eea{\end{eqnarray}}
\def\nn{\nonumber}
\def\beq{\begin{equation}}
\def\eeq{\end{equation}}
\def\ba{\beq\new\begin{array}{c}}
\def\ea{\end{array}\eeq}
\def\be{\ba}
\def\ee{\ea}

\def\tr{{\rm Tr}}

\def\CF{{\cal F}}

\makeatletter
\newdimen\normalarrayskip              
\newdimen\minarrayskip                 
\normalarrayskip\baselineskip
\minarrayskip\jot
\newif\ifold             \oldtrue            \def\new{\oldfalse}
\def\arraymode{\ifold\relax\else\displaystyle\fi} 
\def\eqnumphantom{\phantom{(\theequation)}}     
\def\@arrayskip{\ifold\baselineskip\z@\lineskip\z@
     \else
     \baselineskip\minarrayskip\lineskip2\minarrayskip\fi}
\def\@arrayclassz{\ifcase \@lastchclass \@acolampacol \or
\@ampacol \or \or \or \@addamp \or
   \@acolampacol \or \@firstampfalse \@acol \fi
\edef\@preamble{\@preamble
  \ifcase \@chnum
     \hfil$\relax\arraymode\@sharp$\hfil
     \or $\relax\arraymode\@sharp$\hfil
     \or \hfil$\relax\arraymode\@sharp$\fi}}
\def\@array[#1]#2{\setbox\@arstrutbox=\hbox{\vrule
     height\arraystretch \ht\strutbox
     depth\arraystretch \dp\strutbox
     width\z@}\@mkpream{#2}\edef\@preamble{\halign
\noexpand\@halignto
\bgroup \tabskip\z@ \@arstrut \@preamble \tabskip\z@ \cr}%
\let\@startpbox\@@startpbox \let\@endpbox\@@endpbox
  \if #1t\vtop \else \if#1b\vbox \else \vcenter \fi\fi
  \bgroup \let\par\relax
  \let\@sharp##\let\protect\relax
  \@arrayskip\@preamble}
\def\eqnarray{\stepcounter{equation}%
              \let\@currentlabel=\theequation
              \global\@eqnswtrue
              \global\@eqcnt\z@
              \tabskip\@centering
              \let\\=\@eqncr
              $$%
 \halign to \displaywidth\bgroup
    \eqnumphantom\@eqnsel\hskip\@centering
    $\displaystyle \tabskip\z@ {##}$%
    \global\@eqcnt\@ne \hskip 2\arraycolsep
         $\displaystyle\arraymode{##}$\hfil
    \global\@eqcnt\tw@ \hskip 2\arraycolsep
         $\displaystyle\tabskip\z@{##}$\hfil
         \tabskip\@centering
    &{##}\tabskip\z@\cr}
\begingroup\ifx\undefined\newsymbol \else\def\input#1 {\endgroup}\fi
\input amssym.def \relax
\input amssym
\newfont{\hr}{msbm10}
\newfont{\ams}{msam10}

\font\numbers=cmss12
\font\upright=cmu10 scaled\magstep1
\def\stroke{\vrule height8pt width0.4pt depth-0.1pt}
\def\topfleck{\vrule height8pt width0.5pt depth-5.9pt}
\def\botfleck{\vrule height2pt width0.5pt depth0.1pt}
\def\Zmath{\vcenter{\hbox{\numbers\rlap{\rlap{Z}\kern 0.8pt\topfleck}\kern
2.2pt
                   \rlap Z\kern 6pt\botfleck\kern 1pt}}}
\def\Qmath{\vcenter{\hbox{\upright\rlap{\rlap{Q}\kern
                   3.8pt\stroke}\phantom{Q}}}}
\def\Nmath{\vcenter{\hbox{\upright\rlap{I}\kern 1.7pt N}}}
\def\Cmath{\vcenter{\hbox{\upright\rlap{\rlap{C}\kern
                   3.8pt\stroke}\phantom{C}}}}
\def\Rmath{\vcenter{\hbox{\upright\rlap{I}\kern 1.7pt R}}}
\def\Z{\ifmmode\Zmath\else$\Zmath$\fi}
\def\Q{\ifmmode\Qmath\else$\Qmath$\fi}
\def\N{\ifmmode\Nmath\else$\Nmath$\fi}
\def\C{\ifmmode\Cmath\else$\Cmath$\fi}
\def\R{\ifmmode\Rmath\else$\Rmath$\fi}

\newcounter{app}

\def\app{\setcounter{equation}{0}
\def\theequation{\Alph{app}.\arabic{equation}}\par
   \addvspace{4ex}
   \@afterindentfalse
  \secdef\@app\@dapp}

\newcommand\@app{\@startsection {app}{1}{0ex}%
                                   {-3.5ex \@plus -1ex \@minus -.2ex}%
                                   {2.3ex \@plus.2ex}%
                                   {\normalfont\Large\bf}}
\def\@dapp#1{%
{\parindent \z@ \raggedright  \bf #1}\par\nobreak}
\def\l@app#1#2{\ifnum \c@tocdepth >\z@
    \addpenalty\@secpenalty
    \addvspace{1.0em \@plus\p@}%
    \setlength\@tempdima{8em}%
    \begingroup
      \parindent \z@ \rightskip \@pnumwidth
      \parfillskip -\@pnumwidth
      \leavevmode \bfseries
      \advance\leftskip\@tempdima
      \hskip -\leftskip
      #1\nobreak\hfil \nobreak\hb@xt@\@pnumwidth{\hss #2}\par
    \endgroup\fi}
\newcounter{sapp}[app]

\def\sapp{\def\theequation{\Alph{app}.\arabic{equation}}
\par
\@afterindentfalse
  \secdef\@sapp\@dsapp}
\newcommand{\@sapp}{\@startsection{sapp}{2}{\z@}%
                                     {-3.25ex\@plus -1ex \@minus -.2ex}%
                                     {1.5ex \@plus .2ex}%
                                     {\normalfont\large\bfseries}}

\def\@dsapp#1{%
{\parindent \z@ \raggedright  \bf #1
}\par\nobreak}
\newcommand{\l@sapp}{\@dottedtocline{2}{1.5em}{2.3em}}

\def\2{{1\over 2}}
\def\N2{${\cal N}=2$}

\def\be{ \begin{eqnarray} }
\def\ee{ \end{eqnarray} }

\def\bea{\begin{eqnarray}}
\def\eea{\end{eqnarray}}
\def\nn{\nonumber}

\def\beq{\begin{equation}}
\def\eeq{\end{equation}}
\def\ba{\beq\new\begin{array}{c}}
\def\ea{\end{array}\eeq}
\def\be{\ba}
\def\ee{\ea}

\def\arcsinh{\hbox{ arcsinh}}
\def\arccosh{\hbox{ arccosh}}

\title{Integrability in string/field theories and
Hamiltonian Flows in the Space of Physical Systems}

\author{A.Mironov$^{1,2}$\thanks{mironov@lpi.ac.ru, mironov@itep.ru},
\\ \normalsize \em $^{1}$ Theory
Department, Lebedev Physics Institute, Moscow
~117924, Russia
\\
\normalsize \em $^{2}$ITEP, Moscow
117259, Russia}

\date{}

\begin{document}

\maketitle

\vspace{-8.2cm}

\begin{center}
\hfill FIAN/TD-2002\\
\hfill ITEP/TH-23/02\\
\hfill hep-th/0205202
\end{center}

\vspace{5.5cm}

\begin{abstract}
Integrability in string/field theories is known to emerge when considering
dynamics in the moduli space of physical theories. This implies that one has
to look at the dynamics with respect to unusual time variables like coupling
constants or other quantities parameterizing configuration space of physical
theories. The dynamics given by variations of coupling constants
can be considered as a canonical transformation or,
infinitesimally, a Hamiltonian flow in the space of physical systems. We
briefly consider here an example of mechanical integrable systems. Then,
any function $T(\vec p, \vec q)$ generates
a one-parametric family of integrable systems
in vicinity of a single system.
For integrable system with several coupling constants the
corresponding ``Hamiltonians" $T_i(\vec p, \vec q)$
satisfy Whitham equations and
after quantization (of the original system) become
operators satisfying the zero-curvature condition in
the space of coupling constants:
$$
\left[\frac{\partial}{\partial g_a} - \hat T_a(\hat{\vec p},\hat{\vec q}),\
\frac{\partial}{\partial g_b} - \hat T_b(\hat{\vec p},\hat{\vec q})\right] = 0
$$
\end{abstract}

\section{Introduction}

One of the main lessons we learnt from string theory is that, in contrast
to the usual field theory approach, one should not study the theory at
fixed values of parameters, but instead needs to vary as many parameters as
possible in order to reveal some new structures behind the theory. In fact,
most of the structures realized during last years just can not be discovered
at fixed values of parameters. Moreover, if the theory is restricted by some
additional requirements (which is often the case in the standard field
theory, where one typically asks for renormalizability, unitarity etc.) the
simple structures are typically absent, or become realized in a very non-linear
complicated way. The situation is much similar to the integrable system
which is often can be described by as a simple free system on a manifold but
gets a non-trivial dynamics after the Hamiltonian reduction on the phase
space.

Among new interesting structures realized in such a way is integrability of
complete effective actions in quantum field theory \cite{int}. By complete effective
action we understand the generating function of all correlation functions.
It turns out that the same effective action also enjoys other
interesting features, typically of topological or similar nature.

In this short review, in ss.2 and 3 we are going to list several typical
examples of this kind of structures. However, one should remark that the
set of examples is far not exhausted by the effective actions. One sometimes
needs to deal with some more refined quantities like metrics
of the $\sigma$-model describing the low-energy effective action. This is
the case in Seiberg-Witten theory discussed in ss.4 and 5. In ss.6 and 7 we
develop, following \cite{MM2}, a formalism that could explain all observed
structures of the
effective theories, in particular, we propose some explanation of
integrability of the effective actions. At last, we consider several
manifest examples in ss.8 and 9.

\section{Matrix model as a toy example}

To be more specific, let us consider a toy example of the matrix integral
which imitates path integral for a field theory. This example has come from
the theory of non-critical strings, \cite{GMKBDS}. Namely, we consider
the integral over Hermitian $N\times N$ matrix with the corresponding Haar
measure. The partition function of this theory is given in the simplest case
by the integral
\be\label{1}
Z_N(g)=\int dM e^{\tr M^2+g\tr M^3}
\ee
The first term in the exponential is a counterpart of the kinetic term, while
the second one generates the cubic interaction. Therefore, one can deal with
this integral using the perturbation theory in the parameter $g$. We
restrict ourselves here with maximum the cubic interaction in order to
reproduce situation of the field theory with typically finite number of
different (polynomial) interactions.

Now, if one wants to learn something about this integral, one should deform
the theory to include more interactions and allow constants in integral
(\ref{1}) to vary. Say, let the size of matrix $N$ be a variable, while another
variable $t$ be the coefficient in front of linear term that is put zero in (\ref{1}).
Then, one can easily show \cite{GMMMO} that
the partition function as a function of these two parameters satisfies the
difference-differential equation
\be
{\partial^2\phi_N\over\partial t
^2}=e^{\phi_{N+1}-\phi_{N}}-e^{\phi_N-\phi_{N-1}},\ \ \ \
e^{\phi_N}\equiv{Z_{N+1}\over Z_N}
\ee
One recognizes in this equation the Toda chain equation, $Z_N$ being its
$\tau$-function \cite{Toda}. The Toda chain is an integrable system, therefore, there
are (infinitely) many Poisson-commuting conserved quantities in the system,
each of them can be taken as
a Hamiltonian, i.e. each of them gives rise to a time flow. Since all these
Hamiltonians are Poisson-commuting, we have (infinitely) many commuting time flows
which lead to a whole
hierarchy of equations of motion. This hierarchy is called Toda chain
hierarchy. Note that all the equations of the hierarchy are described by a
single function of (infinitely) many time variables called $\tau$-function.
We expect this $\tau$-function to be the partition function of our system.

Now, what are these other time variables? In fact, they have very simple
meaning: these are nothing but other couplings, the couplings with
$\tr M^k$ \cite{GMMMO}.
Thus, in order to obtain complete integrable system, one needs to consider
all single trace interactions in (\ref{1}), the corresponding coupling
constants being time variables in the Toda chain hierarchy and the partition
function being the Toda chain $\tau$-function. Note that this partition
function is the generating function for all correlators in the theory.

In fact, introducing infinitely many interactions allows one to
construct some other underlying structures in the theory apart from
integrability. Say, if one makes polynomial changes of variable in (\ref{1})
with all single trace interactions included, it leads \cite{MM} to a set of
constraints (Ward identities) satisfied by the partition function:
\be
L_n\left(\int dM e^{\sum_k t_k\tr M^k}\right)=0,\ \ \ \ n\ge -1,
\ \ \ \ L_n\equiv\sum_k
kt_k{\partial\over\partial t_{k+n}}+\sum_{k=0}^n{\partial^2\over\partial
t_{n-k}\partial t_k}
\ee
These constraints form the Borel subalgebra of the Virasoro algebra. Note
that they essentially involve the whole set of times and, therefore, can not
be observed for a truncated ``action'' like (\ref{1}) at all (in contrast to
integrability which we partially observed already with two parameters
included). Indeed, the Ward identities contains more subtle information
about the theory, basically related to the exact (Polchinski) renormalization
group \cite{RG}. This is why their solution typically
unambiguously gives the partition function \cite{FKN}.

\section{Integrability in string/field theories}

Thus, the main lesson we should learn from our simple example is that one
has to switch on and vary as many couplings as possible. Then, for the
complete effective action (generating function of all correlators) one
realizes a series of different underlying structures like integrability,
complete set of Ward identities etc.

Note that the very phenomenon of {\it classical} integrability in different
{\it quantum} systems was observed many times. The examples were mostly done
in quantum integrable systems, just as only these systems
could be exactly solved and
the integrability could be {\it a posteriori} realized. However, the
examples constructed come from different fields (of two-dimensional quantum
field theories, statistical lattice partition functions, etc) \cite{AR}-\cite{Kor}
which implies the phenomenon is very general. Moreover, it was also observed
in matrix models of absolutely different type (so called continuous matrix
models \cite{GKM}), in two-dimensional gauge theories \cite{GKM2U} etc.
These examples are all two-dimensional, however, we mention below
higher-dimensional examples too. Moreover, within the string theory
approach it is also
expected that the sum of the whole perturbative series for the amplitudes
(and the partition function) can be described by a quantum integrable system
\cite{Mor2,int}. Thus, the very fact that the effective
action for the {\it quantum} system is a $\tau$-function of some
{\it classical} integrable system\footnote{It is interesting that all the
examples can be parted into two
large classes of systems naturally depending on times (coupling
constants) and on the so called Miwa variables, these latter ones being
just eigenvalues of an (infinite) external matrix. An instance of the first
class system where the matrix model (\ref{1}) also lies is the quantum non-linear
Schr\"odinger model. The generating functional of the correlators in
this model is a $\tau$-function of some classical integrable
non-linear equation of the same (non-linear Schr\"odinger) kind \cite{BIKS}.
A typical example of the other type
is given by the partition function in the six-vertex model with
non-trivial boundary conditions. This partition function turns out to be the
$\tau$-function of the two-dimensional Toda system expressed in Miwa variables
\cite{Kor}. Another typical representative of the same class of theories is
the continuous matrix models \cite{GKM,GKM2U}. In particular, such matrix models
have much to do with the very low-energy limit of QCD \cite{QCD}.}
is already known for many years,
although has never been explained. In section 6, we propose a way for an
explanation of this fact, following \cite{MM2}.

However, effective actions depends not only on coupling constants. Other
parameters essentially determining the theory are boundary conditions \cite{Morr}
or, which is basically the same, vacuum expectation values of different fields.
It turns out that involving these parameters makes general structures behind the
theory much richer. Strikingly amazing, in this case integrability is also
presented although realized in a different way.

\section{Integrability in Seiberg-Witten theory}

A basic example of the effective action depending on vacuum expectation values
is the low-energy effective action in $N=2$ supersymmetric gauge theories in
the dimensions 4, 5 and 6 with various matter contents. The exact solution
for such an action was constructed by N.Seiberg and E.Witten \cite{SW}, while
its integrable properties were revealed in \cite{GKMMM} (one may find a lot of
discussions of Seiberg-Witten solutions and their integrable structures in the
book \cite{Braden}). More concretely, one studies
the theory with the gauge group $G$ such that the symmetry breaks down to
$U(1)^n$ with
$n$ scalar fields getting non-zero vacuum expectation values, $<\phi_i>=a_i$
and $n$ photons remaining massless (Coulomb branch). One then effectively
obtains at low energies $n$ $N=2$ supersymmetric photodynamics. The scalar
part of the action can be described by a sigma-model with the metric
being the second derivative of a single holomorphic function,
$T_{ij}(\phi)=\partial^2_{i,j}{\cal F}(\phi)$ called prepotential. Singularities of
this function are known from physical arguments (of duality
and compatibility with the renormalization group flows) \cite{SW} and fix ${\cal F}(\phi)$.
The prepotential completely fixes the exact low energy amplitudes in the
theory. It can be constructed as follows.

First of all, one has to find out proper variables whose modular
properties fit the field theory interpretation. These variables are
the integrals of a meromorphic 1-form $dS$
over the cycles on a two-dimensional Riemann surface,
$a_i$ and $a^D_i$
\be\label{aad}
a_{i}=\oint_{A_{i}}dS, \ \ \ \
a_{D}=\oint_{B_{i}}dS,
\ee
(where $i,j=1,....,N_{c}-1$ for the gauge group $SU(N_{c})$).

These integrals play the two-fold role in the Seiberg-Witten approach.
First of all, one may calculate the prepotential ${\cal F}$ and, therefore,
the low
energy effective action through the identification of
$a_D$ and $\partial \CF/\partial a$ with
$a$ defined as a function of moduli (values of condensate) by formula
(\ref{aad}).
Then, using the property of the differential $dS$
that its variations w.r.t. moduli are holomorphic one may also calculate the
matrix of coupling constants in the gauge theory
\be\label{Tij}
T_{ij}(u)=\frac{\partial^{2}{\cal{F}}}{\partial a_{i} \partial a_{j}},
\ee

The second role of formula (\ref{aad}) is that, as
was shown these integrals define the spectrum of the stable states
in the theory which saturate the Bogomolny-Prasad-Sommerfeld (BPS)
limit. For instance, the formula for the BPS spectrum in the $SU(2)$
theory reads as
\be\label{emd}
M_{n,m}=\left|na(u) +ma_{D}(u)\right|,
\ee
where the quantum numbers $n,m$ correspond to the ``electric" and ``magnetic"
states.

It was realized in \cite{GKMMM} that each Seiberg-Witten solution can be
associated with an integrable system with finite number degrees of freedom,
and the prepotential can be immediately constructed from integrable data.

Indeed, the structures underlying Seiberg-Witten
theory are the following set of data \cite{SW}:
\begin{itemize}
\item
Riemann surface ${\cal C}$
\item
moduli space ${\cal M}$ (of the curves ${\cal C}$), the moduli space
of vacua of the gauge theory
\item
meromorphic 1-form $dS$ on ${\cal C}$
\end{itemize}
Exactly this input can be naturally described
within the framework of some underlying integrable system
\cite{GKMMM,Itoyama,Braden}.

To this end, first, we introduce bare spectral curve $E$ that is torus
$y^2=x^3+g_2x^2+g_3$ for the UV-finite
gauge theories with the associated holomorphic 1-form
$d\omega=dx/y$. This bare spectral curve degenerates into the
double-punctured sphere (annulus) for the asymptotically free theories
(where dimensional transmutation occurs): $x\to
w+1/w$, $y\to w-1/w$, $d\omega=dw/w$.
On this bare curve, there are given a
matrix-valued Lax operator $L(x,y)$. The corresponding dressed spectral
curve ${\cal C}$ is defined from the formula $\det(L-\lambda)=0$.

This spectral curve is a
ramified covering of $E$ given by the equation
\be
{\cal P}(\lambda;x,y)=0
\ee
In the case of the gauge group  $G=SU(N_c)$, the function ${\cal P}$ is a
polynomial of degree $N_c$ in $\lambda$.

Thus, we have the spectral curve ${\cal C}$, the moduli space ${\cal M}$ of the
spectral curve being given just  by
coefficients of ${\cal P}$.
The third important ingredient of the construction is the
generating 1-form $dS \cong \lambda d\omega$ meromorphic on
${\cal C}$ (``$\cong$" denotes the equality modulo total derivatives).
{}From the point of view of
the integrable system, it is just the shortened action "$pdq$" along the
non-contractible contours on the Hamiltonian tori. This means that the
variables $a_i$ in (\ref{aad}) are nothing but the action variables in the
integrable system. The defining property of $dS$ is
that its derivatives with respect to the moduli (ramification points)
are holomorphic differentials on the spectral curve.  This, in particular,
means
that
\be
{\partial dS\over\partial a_i}=d\omega_i
\ee
where $d\omega_i$ are the canonical holomorphic differentials\footnote{I.e.
satisfying the conditions
$$
\oint_{A_i}d\omega_j=\delta_{ij},\ \ \ \ \oint_{B_i}d\omega_j=T_{ij}
$$
}.
Integrating this formula over $B$-cycles and using that
$a_D=\partial \CF/\partial a$, one immediately obtains (\ref{Tij}).

So far we reckoned without matter hypermultiplets.
In order to include them, one just needs to consider the
surface ${\cal C}$ with punctures. Then, the hypermultiplet
masses are proportional to the
residues of $dS$ at the punctures, and the moduli space has to be extended to
include these mass moduli. All other formulas remain in essence the same.

The prepotential ${\cal F}$
and other ``physical" quantities are defined in terms of the
cohomology class of $dS$, formula (\ref{aad}). Note that
formula (\ref{Tij}) allows one to identify the prepotential with logarithm of
the
$\tau$-function of the Whitham hierarchy \cite{typeB}: ${\cal F}=\log\tau$.
This illustrates how another, Whitham integrability also emerges in Seiberg-Witten
theories \cite{RGSW}. In fact, this phenomenon is quite general: in matrix models
one observes the Whitham integrability as well \cite{LG}. However, any clear general
reason for this is still missed.

Note that Seiberg-Witten theory celebrates even more interesting properties. One of
them is that the prepotential ${\cal F}$ satisfies a set of highly
non-linear equations called associativity (or WDVV) equations \cite{WDVV}.

\section{Duality in Seiberg-Witten theory}

In order to understand another interesting feature of Seiberg-Witten theory,
one needs to look at the whole web of $N=2$ SUSY gauge theories, in
different space-time dimensions and with different matter contents
\cite{Braden}. Restricting oneself to the theories with one adjoint matter
hypermultiplet and varying the dimensions, one rolls among the members of
the dual Calogero-Ruijsenaars integrable
family \cite{Rui}. For instance, the perturbative limit
of the four-dimensional theory is described by the trigonometric Calogero system which is
dual to a degenerated perturbative limit of the five-dimensional theory
described by the rational Ruijsenaars system. At the same time, the
perturbative limit of the five-dimensional theory is described by the
self-dual trigonometric Ruijsenaars system. Meanwhile, dealing with complete
non-perturbative contributions requires elliptic models\footnote{Note that
the system dual to the complete non-perturbative four-dimensional theory is
the perturbative six-dimensional theory compactified onto two-dimensional
torus. Thus, taking into account non-perturbative (instanton) corrections is
effectively equivalent to summing up Kaluza-Klein modes when compactifying
onto two-dimensional torus.} and ultimately knowledge of
the most general (self-dual) member of the family, the double-elliptic system
\cite{dual,dell}. This system
describes the (non-perturbative) six-dimensional gauge theory compactified
onto two-dimensional torus and is constructed
so far only basing on (self-)duality arguments.

In order to understand what this duality means we develop below the scheme
\cite{MM2}
which simultaneously allows one to understand a reason for quantum systems to
reveal an integrability in coupling constant flows we discussed in sections 2
and 3. Meanwhile, we
realize that the notion of (self-)duality emerged is not very restrictive for the theory,
moreover, there are a lot of self-dual systems. Self-duality of the
Calogero-Ruijsenaars family turns out to have rather more to do with the group
theory interpretation of this family but does not pick up it among other integrable systems.

In order to make the main idea of \cite{MM2} clear,
we start with a simple mechanical
integrable system\footnote{We do not discuss here to what extent we need here the
integrability of the system on the whole phase space. Indeed, since our
consideration is local, one may expect that we need only $N$ Poisson-commuting
Hamiltonians given {\it locally}, at a patch of the phase space.
For the sake of simplicity, we ignore this point here, merely assuming our system
is completely integrable.}.
Integrable system with $N$ coordinates $q_i$ and $N$ momenta $p_i$
is characterized by existence of $N$ Poisson-commuting Hamiltonians
$H_i(\vec p,\vec q)$,
$\{H_i, H_j\} = \frac{\partial H_i}{\partial \vec p}\frac{\partial H_j}{\partial \vec q} -
\frac{\partial H_j}{\partial \vec p}\frac{\partial H_i}{\partial \vec q} = 0$.
For such a system one can consider a canonical transformation,
treating these Hamiltonians as new momenta-like or coordinate-like variables.
From now on, this transformation (of which the infinitesimal version
is a certain Hamiltonian flow) will be the main subject of the paper.

To make the problem precise, let us consider a one-parametric family of
integrable models, parameterized by a single coupling constant $g$ such that
the model is {\it free} when $g=0$.
This means that at $g=0$ the Hamiltonians
$H^{(0)}_i(\vec p) = H_i(\vec p,\vec q| g=0)$ are functions only of
momenta $\vec p$, though, for conventional choices of Hamiltonians in
particular applications, these functions can be non-trivial.
The typical examples are:
$H^{(0)}_k(\vec p) = \sum_{i=0}^N p_i^k$ and
$H^{(0)}_k(\vec p) = \sum_{I: |I|=k} \prod_{i \in I} e^{p_i}$
for $p$-rational and $p$-trigonometric models respectively and some elliptic functions
of $\vec p$ for their elliptic generalizations.

The adequate definition of the new canonical variables
$\vec P_g = \vec P (\vec p,\vec q|g)$ and
$\vec Q_g = \vec Q (\vec p,\vec q|g)$\footnote{
In what follows we often omit the label $g$, implying that the
capital letters $P$ and $Q$ always denote the ``dressed"
momenta and coordinates $P_g$ and $Q_g$.
}
is \cite{dual}
\be
H_i^{(0)}(\vec P) = H_i(\vec p,\vec q|g) \\
\widetilde H_i^{(0)}(\vec q) = \widetilde H_i(\vec Q,\vec P|g)
\label{newP}
\ee
where $\widetilde H_i(\vec p,\vec q|g)$ define the Hamiltonians
of the {\it dual} integrable system. This is the exact definition of what is
the duality in integrable mechanical system, in particular, in Seiberg-Witten
theory. It should not be mixed with
``physical" electro-magnetic duality of this theory \cite{SW} which is the
duality between ``electric" and ``magnetic" states in (\ref{emd}).

Note that in (\ref{newP}) the Hamiltonians depend on the
``dressed" variables $\vec P$ and $\vec Q$ and, moreover, $\vec P$ and
$\vec Q$ are interchanged.
The shape of the dual Hamiltonians is dictated by the requirement
that the new variables $\vec P$ and $\vec Q$ are canonical, i.e.
\be
\sum_i dP_i\wedge dQ_i = \sum_i dp_i\wedge dq_i
\ee
and the Poisson brackets are
\be
\{\ ,\ \} = \sum_i \left(\frac{\partial}{\partial p_i}\otimes
\frac{\partial}{\partial q_i} -
\frac{\partial}{\partial q_i}\otimes \frac{\partial}{\partial p_i}\right) =
\sum_i \left(\frac{\partial}{\partial P_i}\otimes
\frac{\partial}{\partial Q_i} -
\frac{\partial}{\partial Q_i}\otimes \frac{\partial}{\partial P_i}\right)
\ee
In examples of section 9 we shall see that this definition of the dual
Hamiltonians, indeed, makes the trigonometric Calogero system dual to the
rational Ruijsenaars one, trigonometric Ruijsenaars system self-dual etc.
However, this same definition implies that any
integrable system has its dual counterpart, i.e. the dualities between
different members of the Calogero-Ruijsenaars family says nothing special
about these specific integrable systems. This is only their interpretation in group theory terms
that makes the duality of special interest in this case. We, however, do not
address more to this point here, dealing instead with generic properties of
integrable systems.

\section{Integrability and flows in coupling constants: General theory}

Relations (\ref{newP}) define $\vec P_g$ as functions of $\vec p,\vec q$ and
the coupling constant $g$.
Of special interest and importance is the infinitesimal version of this
{\it canonical} transformation, considered as a Hamiltonian flow along the
$g$ direction in the space of coupling constants.
Such transformation is generated by a new Hamiltonian $T(\vec p,\vec q; g)$,
according to the rule:
\be
\frac{\partial \vec P_g}{\partial g} =
\left\{ T(\vec P_g, \vec Q_g| g), \vec P_g \right\} = -\frac{\partial T}{\partial \vec Q_g}, \\
\frac{\partial \vec Q_g}{\partial g} =
\left\{ T(\vec P_g, \vec Q_g| g), \vec Q_g \right\}  = \frac{\partial T}{\partial \vec P_g}
\ee
This new Hamiltonian does not commute with the old ones, but it converts them
into a new set of commuting Hamiltonians.

Moreover, instead of considering a pre-given family of integrable systems,
one can use {\it any} function
$T(\vec p,\vec q| g)$ to define a whole one-parametric family of
integrable systems, though explicit construction of the corresponding
Poisson-commuting Hamiltonians is rarely possible.
Further, a multi-parametric family of integrable systems is similarly
generated by {\it any} collection of functions $T_a(\vec p,\vec q| g)$
satisfying the compatibility (Whitham) equations:
\be
\frac{\partial T_b}{\partial g_a} - \frac{\partial T_a}{\partial g_b} +
\left\{ T_a, T_b \right\} = 0
\label{PC}
\ee
Note that after quantization of the original system
these $T_a(\vec p,\vec q|\{g_b\})$
become a family of operators  $\hat T_a(\hat{\vec p},\hat{\vec q}|\{g_b\})$,
satisfying the zero-curvature condition in the space of coupling constants:
$$
\left[\frac{\partial}{\partial g_a} - \hat T_a(\hat{\vec p},\hat{\vec q}),\
\frac{\partial}{\partial g_b} - \hat T_b(\hat{\vec p},\hat{\vec q})\right] = 0
$$
This can serve as an explanation of emergency of {\it classical} integrability
(=the zero-curvature equations) in the study of
quantum integrable systems we
discussed in the previous sections. In particular, it would be interesting
to deal with the cases described in sections 2 and 3 and
associated to integrable mechanical systems with infinitely many degrees of
freedom along the line of this paper.

The coupling constant variation for the dual integrable system is governed by
the dual Hamiltonian $\widetilde T(\vec p,\vec q|g)$: for
\be
\widetilde H_i^{(0)}(\widetilde{\vec P}) = \widetilde H_i(\vec p,\vec q|g)
\ee
(and
$H_i^{(0)}(\vec q) = H_i(\widetilde{\vec Q},\widetilde{\vec P} |g)$)
we have
\be
\frac{\partial \widetilde{\vec P_g}}{\partial g} =
\left\{ \widetilde T(\widetilde{\vec P_g}, \widetilde{\vec Q_g}| g), \widetilde{\vec P_g} \right\} =
-\frac{\partial \widetilde T}{\partial \widetilde{\vec Q_g}}, \\
\frac{\partial \widetilde{\vec Q_g}}{\partial g} =
\left\{ \widetilde T(\widetilde{\vec P_g}, \widetilde{\vec Q_g}| g), \widetilde{\vec Q_g} \right\}  =
\frac{\partial \widetilde T}{\partial \widetilde{\vec P_g}}
\ee

Relation between $\widetilde T(\vec p,\vec q|g)$ and $T(\vec p,\vec q|g)$
becomes especially simple at any self-dual point
(where $\widetilde H_i(\vec p,\vec q\ | g_{SD}) = H_i(\vec p,\vec q\ | g_{SD})$):
\be
\widetilde T(\vec p,\vec q\ | g_{SD}) = T(\vec q,\vec p\ | g_{SD})
\ee
As immediate consequence, {\it any} symmetric function
$T(\vec p,\vec q|g) = T(\vec q,\vec p|g)$ defines a one-parametric family
of self-dual integrable systems. Therefore, not just the duality but even
the property of self-duality of models from the Calogero-Ruijsenaars family (rational
Calogero, trigonometric Ruijsenaars and double elliptic systems) is in no
way specific. We again should claim that the (self-)duality of this family
is interesting only in the group theory context.

Note that it is not that immediate calculation to restore the Hamiltonian
starting from arbitrarily given $T(\vec p,\vec q|g)$. Say, one can build a
perturbative series in the coupling constant $g$. At the self-dual point
for the system with one degree of freedom one takes the expansion $
T(p,q|g)\equiv \sum_{i=1} T_i(p,q)g^{i-1}$,
$P=H(p,q)=p+\sum_{i=1}\Phi_i(p,q)g^i$ etc and obtains that
$\Phi_1(p,q)=\partial_q T_1(p,q)$, $\Phi_2(p,q)=\partial_q T_2(p,q)+...$.
The ultimate result is that the symmetric part of
$\partial_p\Phi_i(p,q)=\partial^2_{q,p} T_i(p,q)+$ is given independently at any order,
while the antisymmetric parts are fixed by $\Phi_i$'s at lower orders:
\be\label{sdg}
\Phi_1'(p,q)-\Phi_1'(q,p)=0,\\
\Phi_2'(p,q)-\Phi_2'(q,p)=\Phi_1''(p,q)\Phi_1(p,q)-(p\leftrightarrow q),\\
\Phi_3'(p,q)-\Phi_3'(q,p)=\Phi_1''(p,q)\Phi_2(p,q)+\Phi_2''(p,q)\Phi_1(p,q)-
{1\over 2}\Phi_1'''(p,q)\Phi_1^2(p,q)-\\-\Phi_1''(p,q)\Phi_1'(p,q)\Phi_1(p,q)
-(p\leftrightarrow q),\\
\ldots
\ee
where all the derivatives are taken w.r.t. to the first variable. A trivial
solution to these equations (\ref{sdg}) is $\Phi_i(p,q)=p^{n_i+1}q^{n_i}$
with arbitrary $\{n_i\}$. This implies that the Hamiltonian $H(p,q)=pf(pq)$
is self-dual with arbitrary function $f$. Indeed, one easily checks this is
the case (note that for such a Hamiltonian $pq=PQ$). In particular, the
rational Calogero Hamiltonian gets to this class of Hamiltonians.

\section{Generating functions and quantization}

Along with the Hamiltonians $T(\vec P, \vec Q | g)$ one can also consider
the generating functions of canonical transformations in question, like
$S(\vec Q, \vec q|g)$ or its Legendre transform
$F(\vec P, \vec q|g)=\vec P\vec Q - S(\vec Q,\vec q|g)$,
such that
\be
-\vec P = \frac{\partial S}{\partial \vec Q}, \ \
\vec p = \frac{\partial S}{\partial \vec q}
\label{Pp}
\ee
and
\be
T = \frac{\partial S}{\partial g}
\label{PC2}
\ee
Of course, for canonical transformation
\be
\frac{\partial P_i}{\partial q_j} + \frac{\partial p_j}{\partial Q_i} = 0,
\label{symder}
\ee
as implied by (\ref{Pp}),
but one should be more careful when drawing similar conclusion from
(\ref{PC2}): the second $g$ derivatives satisfy eq.(\ref{PC}),
because $g$-derivative
is taken at constant $\vec P_g$ and $\vec Q_g$, which themselves depend on $g$.

Similarly,
\be
-\vec Q = \frac{\partial F}{\partial \vec P}, \ \
\vec p = \frac{\partial F}{\partial \vec q}
\ee
At self-dual points $F(\vec P,\vec q | g_{SD}) = F(\vec q,\vec P |
g_{SD})$ is a symmetric function.

Note that
\be
{\partial F\over\partial g}=T
\ee
(with $p$ and $Q$ constant), i.e. $T$ is, in a sense, a more invariant
quantity than $S$ and $F$, which does not depend on the choice of
independent variables.

Another way to see it is to consider a flow from an integrable system with
coupling constant $g_1$ to the same system with coupling constant $g_2$.
Then, $T$ depends only on $g_2$, but not on $g_1$, while the generating
functions depend on both $g_1$ and $g_2$.

Basically, if reexpressed in terms of $\vec Q$ and $\vec q$ (instead of $\vec P$
and $\vec Q$) the Hamiltonian $T = \partial S/\partial g$. Therefore, $e^{iS}$ can be
considered as a kind
of an evolution operator (kernel) in the space of coupling constants,
which performs a canonical transformation from the
free system to the integrable one.
After quantization it can be symbolically represented as
\be
e^{i\hat S}(\vec Q,\vec q|g) = \oplus_{\lambda} |\psi^{(0)}_\lambda(\vec Q)\rangle
c_{\vec \lambda} \langle \psi_\lambda(\vec q)|
\label{evo}
\ee
where $|\psi_\lambda\rangle$ and $|\psi^{(0)}_\lambda\rangle$
are eigenfunctions of the system and $c_{\vec \lambda}$ are some
coefficients, depending on the spectral parameter $\vec \lambda$.
The evolution operator satisfies the quantum version of eq.(\ref{newP}),
\be
H^{(0)}(\partial/\partial \vec Q) =
e^{-i\hat S(\vec Q,\vec q|g)}
H(\partial/\partial \vec q,\vec q|g) e^{\hat iS(\vec Q,\vec q|g)}
\ee
Since eigenfunctions $|\psi^{(0)}_\lambda(Q)\rangle$ of a free system
are just exponents of the spectral parameter $\vec\lambda$,
the dressed eigenfunctions $|\psi_\lambda(q)\rangle$ are basically
Fourier transforms of the evolution operator $e^{iS(Q,q)}$:
\be
\psi_\lambda(q) \sim \int e^{i\hat S(Q,q)} e^{i\lambda Q} dQ
\label{efeS}
\ee
Similarly,
\be
\psi_\lambda(q) \sim e^{i\hat F(P,q)} \delta(\lambda - P) dP \sim e^{i\hat
F(\lambda,q)}
\label{efeF}
\ee
Note that there is a freedom in solutions of eq.(2) to shift $\{Q_i\}$ by an
arbitrary function of $\{P_i\}$. This shift is quite complicated in terms of
the generating function $S$, but in $F$ it is just an addition of the term
depending only on $\{P_i\}$. In the quantum case, this ambiguity in the
definition of $F$ is just a matter of normalization of the eigenfunction.

One can also consider the quantum counterpart of $T(p,q|g)$ which is a
Hamiltonian that gives rise to a Schr\"odinger equation w.r.t. the coupling
constant
\be
{\partial\psi\over\partial g}=i\hat T\psi
\ee
Therefore, the wavefunction $\psi$ can be also realized as a path integral
over the phase space variables $P(g),Q(g)$.

The generating functions $S(Q,q)$, $F(P,q)$ and $T(p,q)$ satisfy the quasiclassical
versions of these relations (when multiple derivatives of $S$, $F$ and $T$ are
neglected).
Exact definition of quantum evolution operators, including effects of
discrete spectra and identification of the spectra at different values of
coupling constants, i.e. precise definition of the spectral parameter
$\lambda$ in such a way that eq.(\ref{evo}) is diagonal in $\lambda$,
will be considered elsewhere.

\section{Particular examples}

In order to illustrate our consideration of the previous sections, we
briefly discuss here several manifest examples (further details can be found
in \cite{MM2}) restricting ourselves to the systems with one degree of
freedom only. We start with the simplest example of harmonic oscillator. Then,
$H(p,q|\omega) = \frac{1}{2}(p^2 + \omega^2q^2)$.
Let the frequency $\omega$ play the role of the coupling constant so that
$H^{(0)}(p) = \frac{1}{2}p^2$. Then
\be
P = \sqrt{p^2 + \omega^2q^2}, \\
Q = \frac{\sqrt{p^2 + \omega^2q^2}}{\omega}\arctan \frac{\omega q}{p}
\ee
The inverse transformation looks like
\be
p = P\cos \frac{\omega Q}{P}, \\
q = \frac{P}{\omega} \sin \frac{\omega Q}{P}
\ee
i.e. the Hamiltonian of the dual flow is
\be
\widetilde H(\widetilde p,\widetilde q|\omega) =
\frac{\widetilde q}{\omega}\sin \frac{\omega \widetilde p}{\widetilde q}
\ee
(Note that with this definition $\widetilde H^{(0)}(p) = p$ and also note that
$p,q$-duality does not respect conventional dimensions of $p$ and $q$ so
that it should not be a surprise that $\omega q/p$ in $H$ got substituted
by $\omega \widetilde p/\widetilde q$ in $\widetilde H$.
Of course, $\widetilde p$ and $\widetilde q$ are nothing but $Q_\omega$ and $P_\omega$.)

The generator of $\omega$-evolution is
\be
T = \frac{P^2}{4\omega^2}\sin\frac{2\omega Q}{P} -
\frac{PQ}{2\omega} = \frac{pq - PQ}{2\omega}
\label{Tho}
\ee
For the dual system we similarly have
\be
\widetilde T = \frac{-\omega\widetilde P\widetilde Q +
(\widetilde Q^2 + \omega^2\widetilde P^2)\arctan\frac{\omega \widetilde P}{\widetilde Q}
}{2\omega^2}
= \nn \\ =
\frac{\widetilde p\widetilde q - \widetilde P\widetilde Q}{2\omega} =
\frac{PQ -
pq}{2\omega} = -T
\ee
The generating function
\be
F(P,q|\omega) = \int pdq = \int\sqrt{P^2 - \omega^2q^2} dq
= \nn \\ =
\frac{1}{2}q\sqrt{P^2 - \omega^2q^2} +
\frac{P^2}{2i\omega} \log\left(i\frac{\omega q}{P} + \sqrt{1-\frac{\omega^2
q^2}{P^2}}\right)
\ee
so that the relation (\ref{efeF}) in quasiclassical approximation is
\be
e^{iF(\lambda, q)} =
\left(i\frac{\omega q}{\lambda} - \sqrt{1-
\frac{\omega^2q^2}{\lambda^2}}\right)^{\lambda^2/2\omega}
e^{{iq\over 2}\sqrt{\lambda^2 - \omega^2q^2}}
\sim \nn \\ \sim
e^{-\omega q^2/2} He_\nu(\sqrt{\omega}q)
\label{efeFho}
\ee
where
$\frac{\lambda^2}{2\omega} = \nu + \frac{1}{2}$
and the Hermite polynomials, satisfying
\be
(-\partial^2_x + x^2)e^{-x^2/2}He_\nu(x) = (2\nu+1)e^{-x^2/2} He_\nu(x)
\ee
are given by inverse Laplace transform of the generating function $e^{-x^2/4 +
t/x}$
\be
He_\nu(x) \sim \int \frac{dt}{t^{\nu + 1}}e^{-t^2/4 + tx}
\ee
Taking this integral in the saddle point approximation, one gets
\be
He_\nu(x) \sim \left(x-\sqrt{x^2-2(\nu+1)}\right)^{\nu+1}
e^{x^2/2+x/2\sqrt{x^2-2(\nu+1)}}
\ee
which, at large $\nu$, is in accordance with eq.(\ref{efeFho}).

For a single particle (one degree of freedom) any dynamics is integrable,
and any Hamiltonian is canonically equivalent to the free one.
Therefore, it may make sense to look at the general Hamiltonian.
For $H = \sqrt{p^2 + g^2V(q)}$ the dual Hamiltonian is $q=\widetilde H(P,Q)$.
The canonicity condition then gives
\be
{\partial \widetilde H(P,Q)\over\partial Q}={p(P,Q)\over P}
\ee
Therefore, the dual Hamiltonian can be obtained via solving the equation
\be\label{q}
Q=\int^q {d\xi\over\sqrt{1-g^2V(\xi)/P^2}}
\ee
w.r.t. $q$. Now taking the derivative of $P=\sqrt{p^2+g^2V(q)}$ w.r.t. $g$,
one obtains
\be\label{dT}
-{\partial T\over\partial Q}={g\over P}V(q(P,Q))
\ee
where $q(P,Q)$ is given by (\ref{q}).

\section{Calogero-Ruijsenaars family}

To complete this our review, we also list some results for the
rational and trigonometric members of the Calogero-Ruijsenaars family (which
describe different limits of Seiberg-Witten theory with adjoint matter
hypermultiplet in different space-time dimensions):

\begin{itemize}

\item The rational-rational case (rational Calogero model)

In this case the single Hamiltonian is
$H = \frac{1}{2}\left(p^2 + \frac{g^2}{q^2}\right)$, thus $H^0(p) = \frac{1}{2}p^2$,
and
\be
P^2 = p^2 - \frac{g^2}{q^2}, \nn \\
q^2 = Q^2 - \frac{g^2}{P^2}
\ee
The rational Calogero model is self-dual. The flow is generated by the Hamiltonian
\be
T(p,q;g) = \frac{1}{2}\log \frac{pq - g}{pq + g} = \frac{1}{2}\log \frac{PQ - g}{PQ + g}
\ee
This is obviously a symmetric function of $p$ and $q$, thus,
$T_D(p,q) = T(p,q)$ as it should be for the self-dual system. The generating
function is
\be
S = g\arccosh\frac{Q}{q} = g\log\left(\frac{Q - \sqrt{Q^2-q^2}}{q}\right) =
\frac{g}{2}\log \frac{Q - \sqrt{Q^2-q^2}}{Q + \sqrt{Q^2-q^2}}, \nn \\
\frac{\partial S(Q,q|g)}{\partial g} = T(p(Q,q),q|g)
\ee
In this case, $S$ is a simple linear function of $g$, and
$\partial S/\partial g = S/g$. Similarly,
\be
F(P,q)=\sqrt{P^2q^2+g^2}+{g\over 2}\log {\sqrt{P^2q^2+g^2}-g\over
\sqrt{P^2q^2+g^2}+g},\\
\frac{\partial F(P,q|g)}{\partial g} = T(p(P,q),q|g)
\ee
In accordance with (\ref{efeS})
\be\label{qBes}
\psi_\lambda(q) \sim \int e^{iS(q,Q|g)} e^{iQ\lambda} dQ \sim qJ_{ig}(i\lambda q)
\ee
where $J_g(x)$ is the Bessel function. The exact quantum wavefunction, solving the
equation $(-\partial^2_x + \frac{g^2}{x^2})  \psi_\lambda(x)=
\lambda^2\psi_\lambda(x)$, is $\sqrt{x}J_\nu(i\lambda x)$, $\nu^2=-g^2+1/4$.
It coincides with (\ref{qBes}) in quasiclassical approximation.
\item The rational-trigonometric case (trigonometric Calogero model)
\be
P^2=2 H= p^2 - \frac{g^2}{\sinh^2 q}, \\
\cosh^2 q =\widetilde H^2=\cosh^2 Q\left(1 - \frac{g^2}{P^2}\right), \\
T = \frac{1}{2}\log \frac{P\tanh Q - g}{P\tanh Q + g}
= \frac{1}{2}\log \frac{p\tanh q - g}{p\tanh q + g}
\ee
In this case $S$ is still a simple linear function of $g$
\be
S=g\arcsinh\left({\sinh Q\over\sinh q}\right)
\ee

\item The trigonometric-rational case (rational Ruijsenaars model)
\be
\cosh^2 P =H^2=\cosh^2 p\left(1 - \frac{\sinh^2\epsilon}{q^2}\right), \\
q^2 = 2\widetilde H=Q^2 - \frac{\sinh^2\epsilon}{\sinh^2 P}, \\
T = \frac{1}{2}\log \frac{Q\tanh P - \tanh\epsilon}{Q\tanh P + \sinh\epsilon}
= \frac{1}{2}\log \frac{q\tanh p - \tanh\epsilon}{q\tanh p + \sinh\epsilon}
\ee
In this case, $S$ is no longer a simple linear function of the coupling constant $\epsilon$.

\item The trigonometric-trigonometric case (trigonometric Ruijsenaars model)
\be
\cosh^2 P =H^2=\cosh^2 p\left(1 - \frac{\sinh^2\epsilon}{\sinh^2 q}\right), \\
\cosh^2 q =\widetilde H^2=\cosh^2 Q\left(1 - \frac{\sinh^2\epsilon}{\sinh^2P}\right), \\
T = \frac{1}{2}\log \frac{\tanh P\tanh Q - \tanh\epsilon}{\tanh P\tanh Q + \tanh\epsilon}
= \frac{1}{2}\log \frac{\tanh p\tanh q - \tanh\epsilon}{\tanh p\tanh q + \tanh\epsilon}
\ee

\end{itemize}

\section{Acknowledgements}

I am grateful to A.Morozov for valuable discussions and to S.Kharchev for reading
the manuscript. The work is partly supported
by grants INTAS 00-561, RFBR-01-02-17682a, the Grant of Support for the Scientific
Schools 96-15-96798 and by the Volkswagen-Stiftung.

\end{document}